 \documentclass[preprint,review,12pt]{elsarticle}

\usepackage{graphicx}
\usepackage{float} 
\usepackage{amssymb}
\usepackage{url, hyperref}
\usepackage{rotating}

\usepackage{lineno}
\usepackage{xcolor}  
\usepackage{siunitx} 
\usepackage{svrsymbols} 
\usepackage[utf8]{inputenc} 
\usepackage[T1]{fontenc} 

\begin{document}

\begin{frontmatter}


\title{Spectroscopic performance of Low-Gain Avalanche Diodes for different types of radiation}



\author[mainaddress]{Gabriele Giacomini\corref{correspondingauthor}}
\author[mainaddress]{Wei Chen}
\author[physics]{Gabriele D'Amen}
\author[mainaddress]{Enrico Rossi}
\author[physics]{Alessandro Tricoli}


\cortext[correspondingauthor]{Corresponding author}

\address[mainaddress]{Instrumentation Division, Brookhaven National Laboratory, Upton, 11973 New York, United States}
\address[physics]{Physics Department, Brookhaven National Laboratory, Upton, 11973 New York, United States}

\begin{abstract}
Low-Gain Avalanche Diodes are a type of silicon Avalanche Photo-Diodes originally developed for the fast detection of minimum ionizing particles in high-energy physics experiments. Thanks to their fast timing performance, the Low-Gain Avalanche Diode paradigm enables detectors to accurately measure minimum ionizing particles with a timing resolution of a few tens of picoseconds. Such a performance is due to a thin substrate and the presence of a moderate signal gain. This internal gain of a few tens is enough to compensate for the reduced charge deposition in the thinner substrate and the noise of fast read-out systems. While Low-Gain Avalanche Diodes are optimized for the detection of minimum ionizing particles for high-energy particle detectors, it is   critical to study their  performance for the detection of different types of particle, such as X-rays, gamma-rays, or alphas.  In this paper, we evaluate the gain of three types of Low-Gain Avalanche Diodes: two devices with different geometries and doping profiles fabricated by  Brookhaven National Laboratory,  and one  fabricated by Hamamatsu Photonics with a different process. 

Since the gain in LGADs depends on the bias voltage applied to the sensor, pulse-height spectra have been acquired for bias voltages spanning from the depletion voltage up to breakdown voltage. The signal-to-noise ratio of the generated signals and the shape of their spectra allow us to probe the underlying physics of the multiplication process.
\end{abstract}


\end{frontmatter}


\section{Introduction}
\label{Introduction}
 Low-Gain Avalanche Diodes (LGADs)~\cite{PELLEGRINI201412} differ from regular Avalanche Photo-Diodes (APDs) as they are optimized for the detection of minimum ionizing particles (mips). Their development originated from the need for precisely measuring the timing of mips in High-Energy Physics (HEP) experiments at the Large Hadron Collider (LHC) to disentangle the effect of multiple interactions per bunch crossing. Arrays of LGADs will be the sensors of choice for the timing sub-detectors of both the CMS \cite{CMS:2667167} and the ATLAS \cite{CERN-LHCC-2020-007} experiments. 
 
 The intrinsic timing resolution of LGADs, in the order of a few tens of picoseconds, is achieved thanks to a combination of 1) the sensors being fabricated on thin substrates (20 - 50~$\mu$m), and 2) the signals being internally amplified with gain in the order of 5 - 100, considerably smaller than that featured in standard APDs. Such gain compensates for the small signal in the thin substrate, makes the jitter contribution to noise small, and allows for a large signal-to-noise ratio even when the sensor is coupled to fast read-out electronics~\cite{HELLER2021165828}.
 
 The gain has however been accurately measured only with mips, which create about 80 electron-hole pairs per micron along a track in the traversed silicon substrate. The mechanism by which other types of radiation produce charge clouds differs from the linear track produced by mips, as the energy deposition can be  more or less deep and localized in silicon and produces a distinct charge density. The gain experienced by different types of radiation needs therefore to be assessed by dedicated measurements, for instance using a low-noise read-out such as a charge sensitive pre-amplifier (CSA).

 In this work, we study the gain of three different types of LGADs  produced in two different batches by Brookhaven National Laboratory (BNL) and Hamamatsu Photonics (HPK) in a different batch.  All LGADs were exposed to X/gamma-rays of different energies, as well as beta (close to mips) and alpha particles.  Measurements were performed at different bias voltages to determine the gain and the excess noise as a function of this parameter.
 
 This paper is organized as follows: in Section 2, we describe the setup, the LGAD devices under test, and the employed radioactive sources; in Section 3, we present the spectroscopic performance of the LGAD extracted from the pulse height spectra, discussing the gain of all sensors in the various scenarios and the extrapolated shape of the distribution of charge deposited by radiation.  

\section{Experimental setup}
\label{The setup}

\subsection{Sensors under test}
\begin{figure}[!htbp]
    \centering\includegraphics[trim={0 0 400 0}, clip, width=0.85\textwidth]{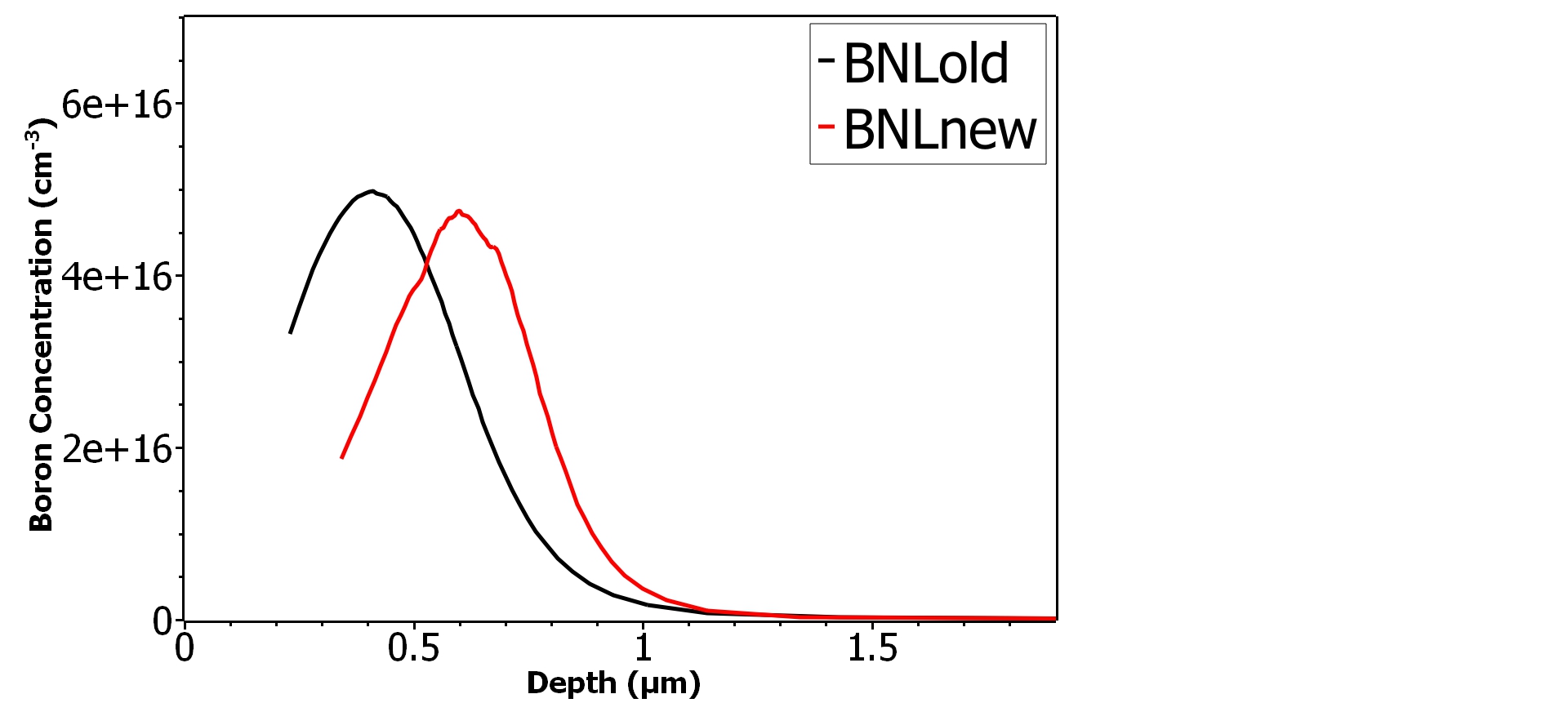}
    \caption{Doping profile of the boron gain layer as extracted from measurements of capacitance as a function of voltage for the two BNL LGADs under test. The position at $x=0$ represents the position of the junction between the n+ layer and the gain layer. Only BNL-fabricated devices are shown.} 
    \label{fig:doping}
\end{figure}

A total of three LGAD devices were tested:
\begin{figure}[!tbp]
    \centering
    \includegraphics[width=0.95\textwidth]{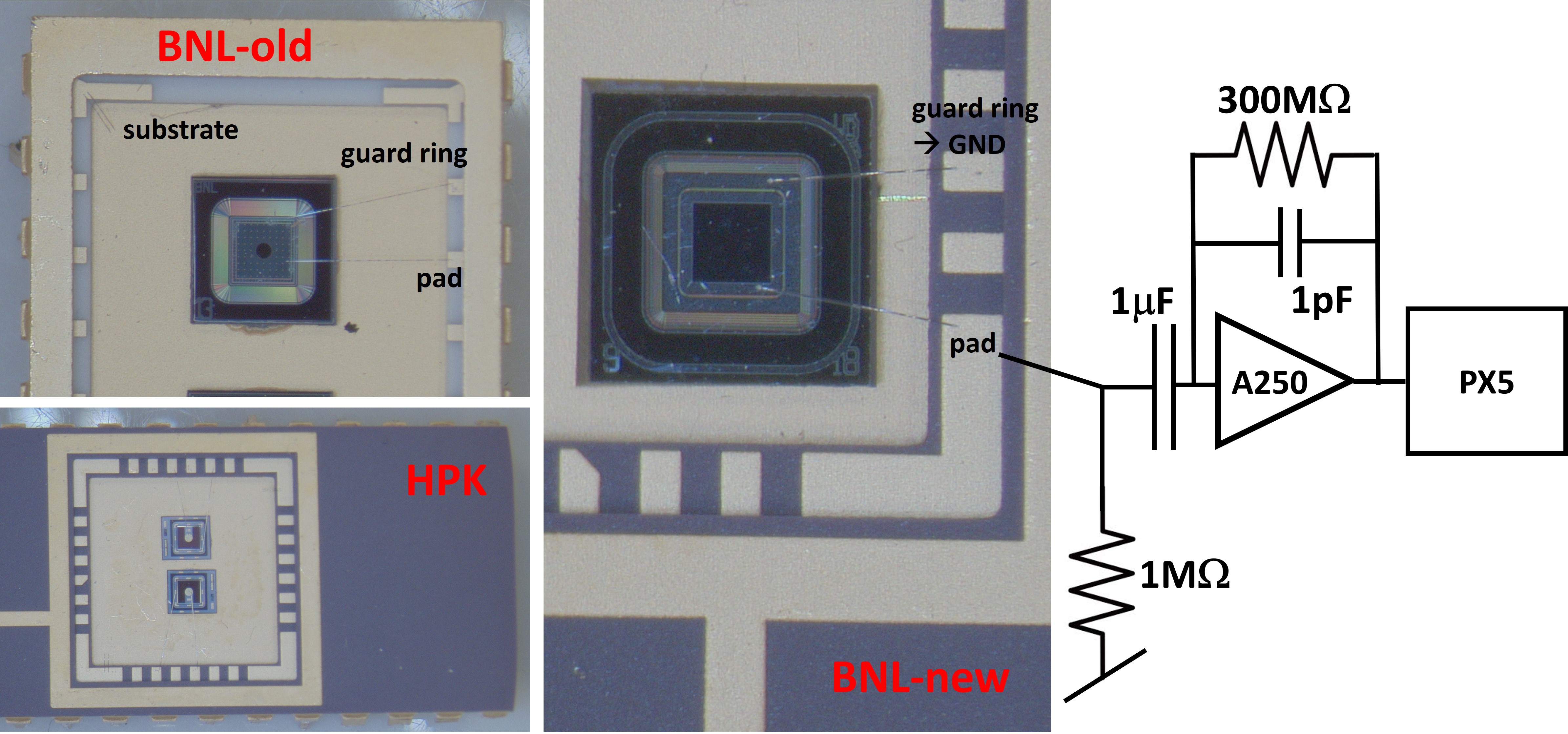}
      \caption{The three devices under test: \textit{top left)} the \textsc{BNL-old} LGAD, \textit{center)} the \textsc{BNL-new} LGAD, and \textit{bottom left)} the \textsc{HPK} LGAD. The schematic \textit{(right)} represents the readout system connected to the sensors as detailed in text body. }
    \label{fig:lgads}
\end{figure}
\begin{itemize}
    \item[$\bullet$] An LGAD sensor, produced as part of the very first BNL fabrication in 2017~\cite{GIACOMINI201952}, has a depletion voltage of 150~V (due to the relatively low resistivity of the epitaxial layer), an active thickness of 50~$\mu$m, a breakdown voltage of 300~V, and an active area of 2$\times$2~mm$^2$. It will be referred to as "\textsc{BNL-old}" in the following.
    \item[$\bullet$] A second LGAD comes from a more recent production by BNL carried out in 2021, it has a total depletion voltage of 30~V, an active thickness of 50~$\mu$m, a breakdown voltage of 200~V, and an active area of 1.3$\times$1.3~mm$^2$ to match the LGAD pixels designed for high-luminosity upgrades of the ATLAS and CMS experiments. It will be referred to as "\textsc{BNL-new}" in the following.
    \item[$\bullet$] The third device under test is an LGAD fabricated by HPK, with a total depletion voltage of 50~V, an active thickness of 35~$\mu$m, a breakdown voltage of 260~V and an active area of 1.3$\times$1.3~mm$^2$. It will be referred to as "\textsc{HPK}" in the following. 
\end{itemize} 
The three devices are shown in Fig.~\ref{fig:lgads}. A major difference among the three LGADs is the depth of the gain layer, obtained via boron implantation: in the \textsc{BNL-old} sensor the gain layer is shallower than the one in \textsc{BNL-new}, with a depth of $\sim$400~nm for the former and $\sim$650~nm for the latter, as shown in Fig.~\ref{fig:doping}; the gain layer of the \textsc{HPK} LGAD is deeper than the one implanted in either BNL devices.
%

\subsection{Readout system}

%
The setup used to characterize the spectroscopic performance of the three LGADs under test is shown in Fig.~\ref{fig:lgads}. 
The sensors are glued on a dual-in-line ceramic board which plugs into a general-purpose breadboard. From here, short wires connect the substrate and the guard ring to the the high voltage and ground lines, respectively. A third wire connects the LGAD pad to the input of the readout system.

The signal pad of the LGAD is AC-coupled to the input of an Amptek A250 low-noise CSA~\footnote{https://www.amptek.com/-/media/ametekamptek/documents/resources/products/specs/a250-specs.pdf} through a large-value capacitance. The leakage current is discharged to ground through a 1~MOhm resistor. The AC-coupling is rendered necessary due to the sensor leakage current, whose magnitude becomes very high at higher bias voltages (when the gain becomes large). Without such AC-coupling scheme, the leakage current  would flow through a 300~MOhm feedback resistor and saturate the CSA.

The output of the A250 CSA is directly fed to an Amptek PX5 Digital Pulse Processor~\footnote{https://www.amptek.com/-/media/ametekamptek/documents/resources/products/specs/px5-spec.pdf} which provides a digital trapezoidal shaping with adjustable peaking times and flat-top widths. For all the measurements, the minimum allowed shaping time of 0.8~$\mu$s has been used, together with a 0.2~$\mu$s flat top width. The PX5 also provides the measurement of the pulse height of acquired signals by means of an internal multi-channel analyzer (MCA). The measured pulse-height spectrum is then plotted in real-time on a PC and stored for off-line analysis. The calibration of the CSA for gain calculation is made by performing the same measurements using a BNL-made 2~mm$\times$2~mm, 500~$\mu$m thick silicon PIN diode, which does not feature any gain.

\subsection{Radiation sources}
Four spontaneous radioactive sources have been used to generate different types of radiation:
\textit{a)} a ${}^{90}$Sr source, emitting beta particles with energies up to 2~MeV that deposit energies close to those of mips; \textit{b)} a ${}^{55}$Fe source of 5.9~keV X-rays; \textit{c)} an ${}^{241}$Am source of gamma-rays up to 60~keV and \textit{d)} a ${}^{241}$Am source of alpha particles with energies up to 5.46~MeV.

These radioactive sources, placed about a centimeter away from the device under test, illuminate the junction side of the devices. 
The entire setup was enclosed in a metallic light-proof box for electromagnetic shielding.

\section{Results}
\begin{figure}[!htbp]
    \centering\includegraphics[width=0.8\linewidth]{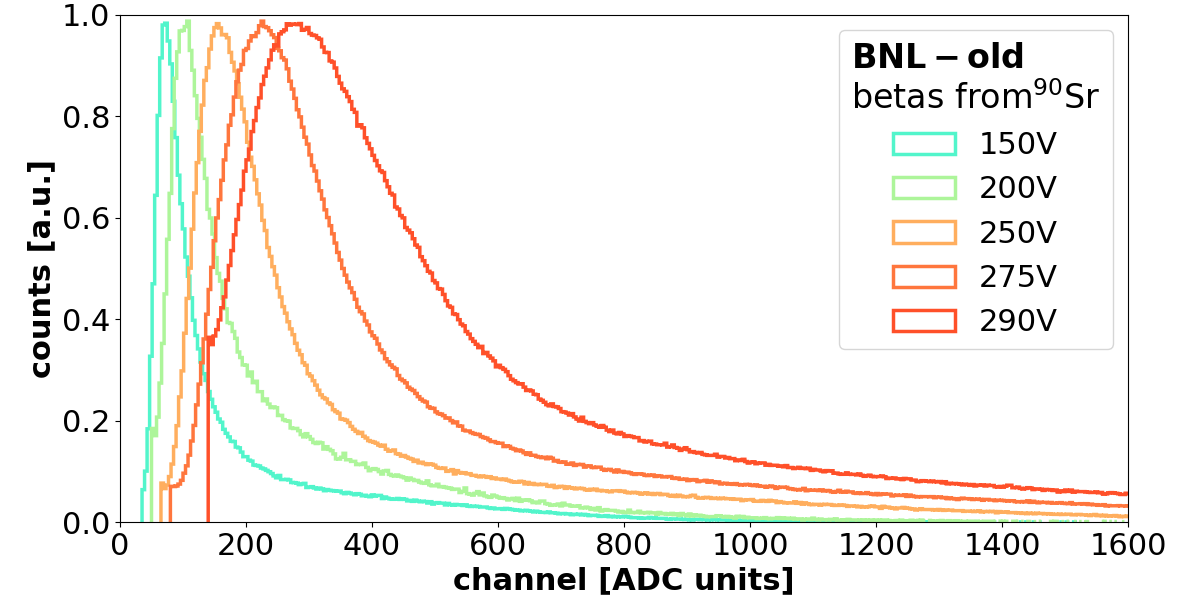}
    \centering\includegraphics[width=0.8\linewidth]{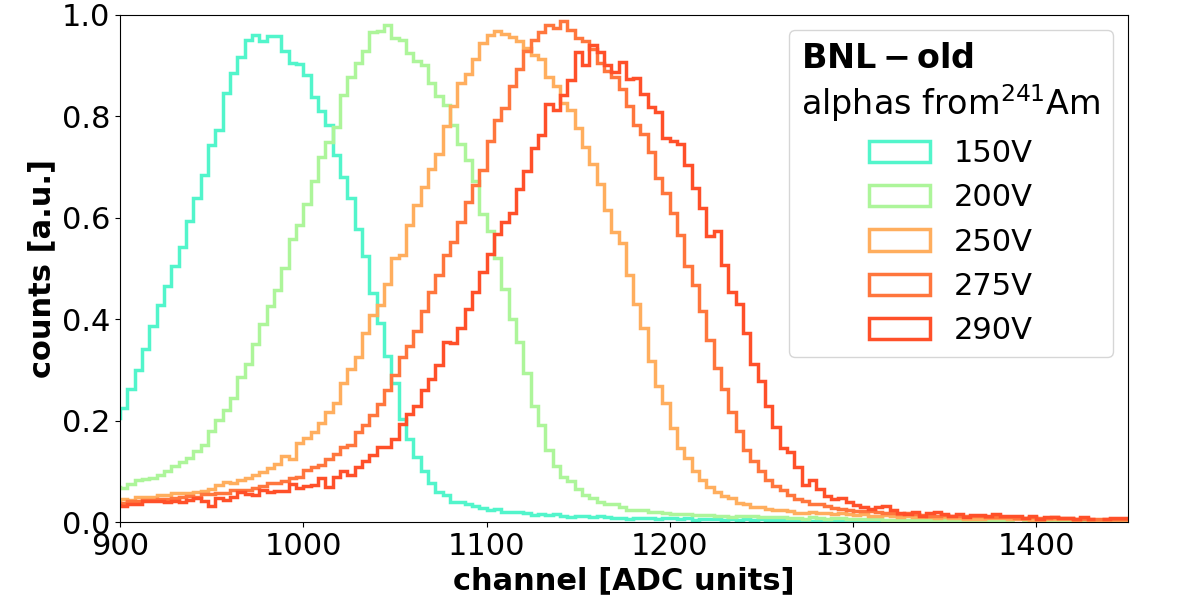}
    \caption{Pulse height spectra acquired by the BNL-old LGAD when exposed to \textit{top)} betas from ${}^{90}$Sr and \textit{bottom)} alphas from ${}^{241}$Am. Counts are normalized for better comparison of shapes. The charge on the \textit{x}-axis is represented in units of ADC "channels". The distributions are cut at lower ADC values to remove the contribution from the sensor noise.}
    \label{fig:BNLold}
\end{figure}

\begin{figure}[!htbp]
    \centering\includegraphics[width=0.8\linewidth]{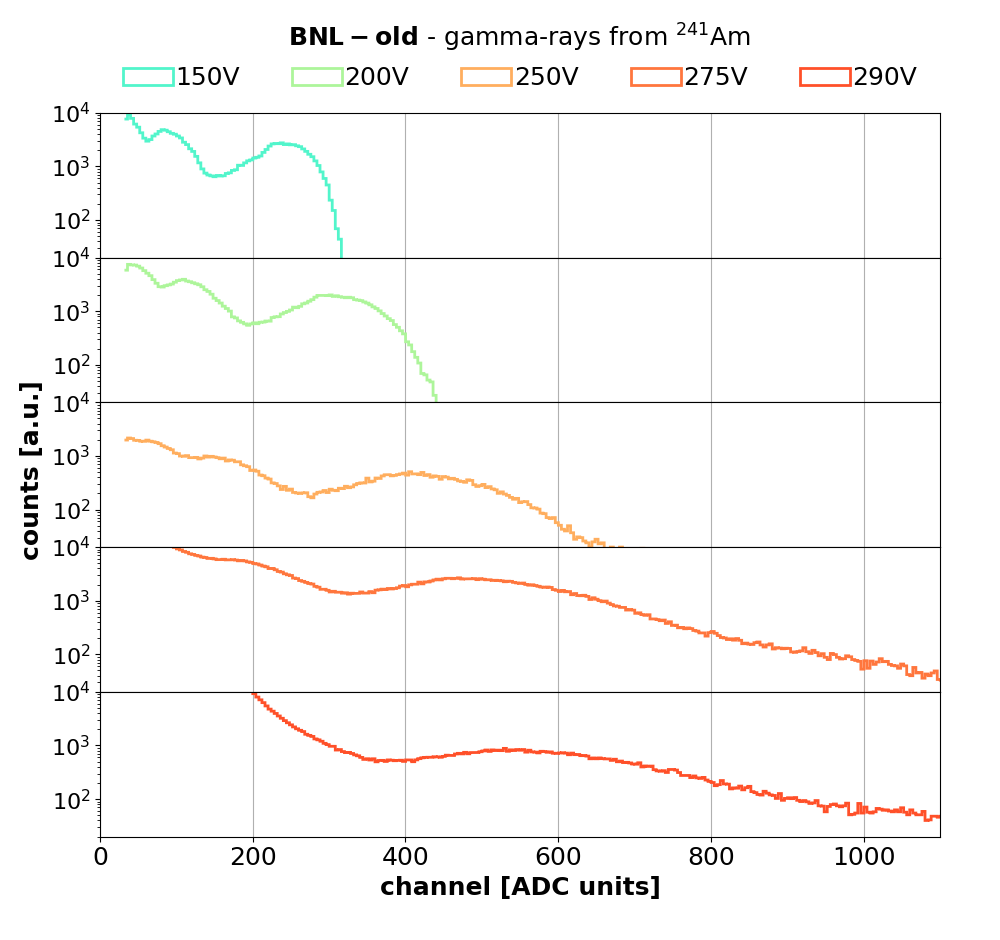}
    \caption{Pulse height spectra acquired by the BNL-old LGAD when exposed to gamma-rays from ${}^{241}$Am. Counts are normalized for better comparison of the shapes. The charge on the \textit{x}-axis is represented in units of ADC "channels". The distributions are cut at lower ADC values to remove the contribution from the sensor noise.}
    \label{fig:BNLold-2}
\end{figure}

Signal spectra generated in LGADs by particles emitted by the four sources were acquired for the three LGADs under study.
Figure~\ref{fig:BNLold} shows the spectra obtained for ${}^{90}$Sr beta particles and ${}^{241}$Am alpha particles acquired with the \textsc{BNL-old} LGAD, while Figure~\ref{fig:BNLold-2} shows the spectra obtained obtained for ${}^{241}$Am gamma-rays. Spectra are acquired for several bias voltages, from depletion (150~V) to the onset of breakdown (290~V). Spectra from ${}^{55}$Fe are not shown for the \textsc{BNL-old} device as the signals were too small compared to the high noise and the limited amplification of the device. All the spectra but the alpha particle one have been acquired by setting the PX5 digital gain to 5; for the alpha particle spectra, as the generated charge is larger, a gain of 0.75 was used.

\begin{figure}[!htbp]
    \centering\includegraphics[width=0.8\linewidth]{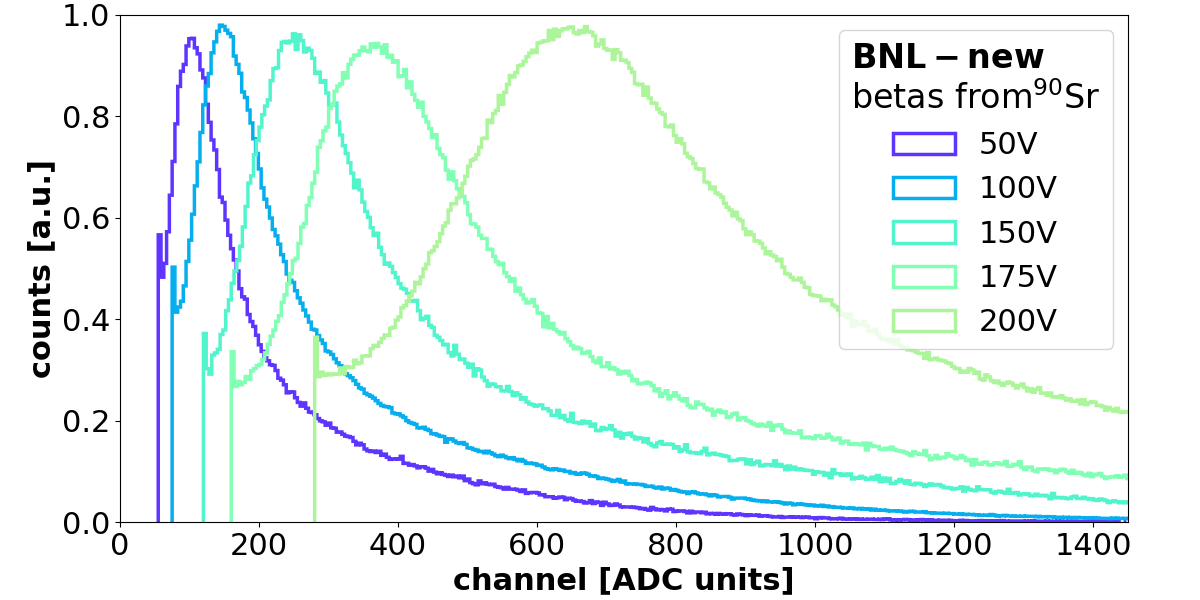}
    \centering\includegraphics[width=0.8\linewidth]{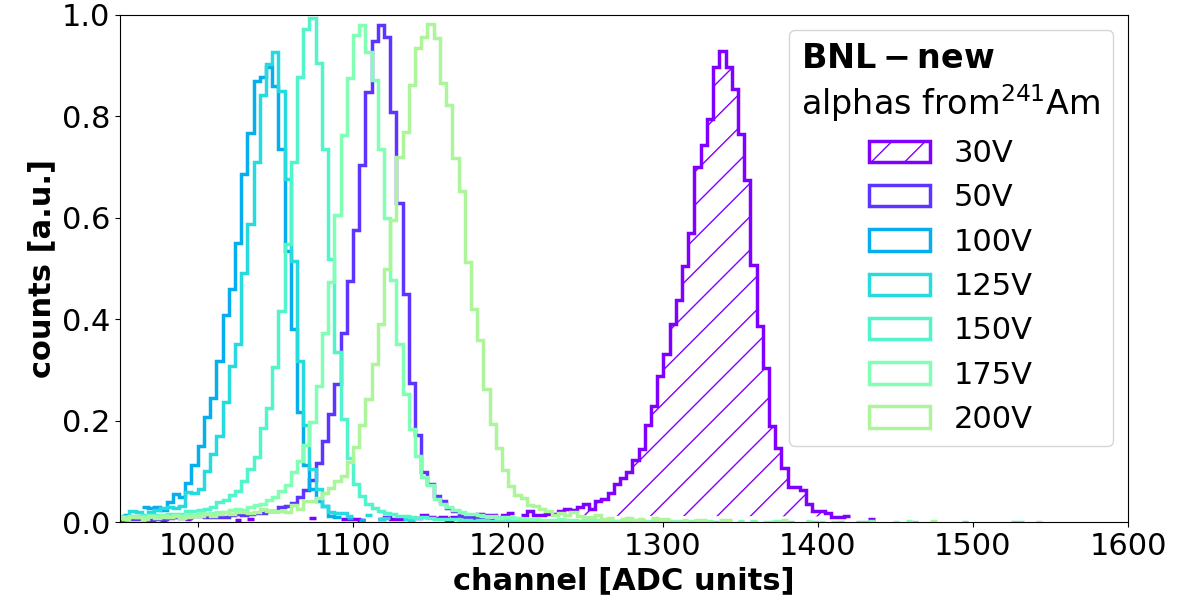}
    \centering\includegraphics[width=0.8\linewidth]{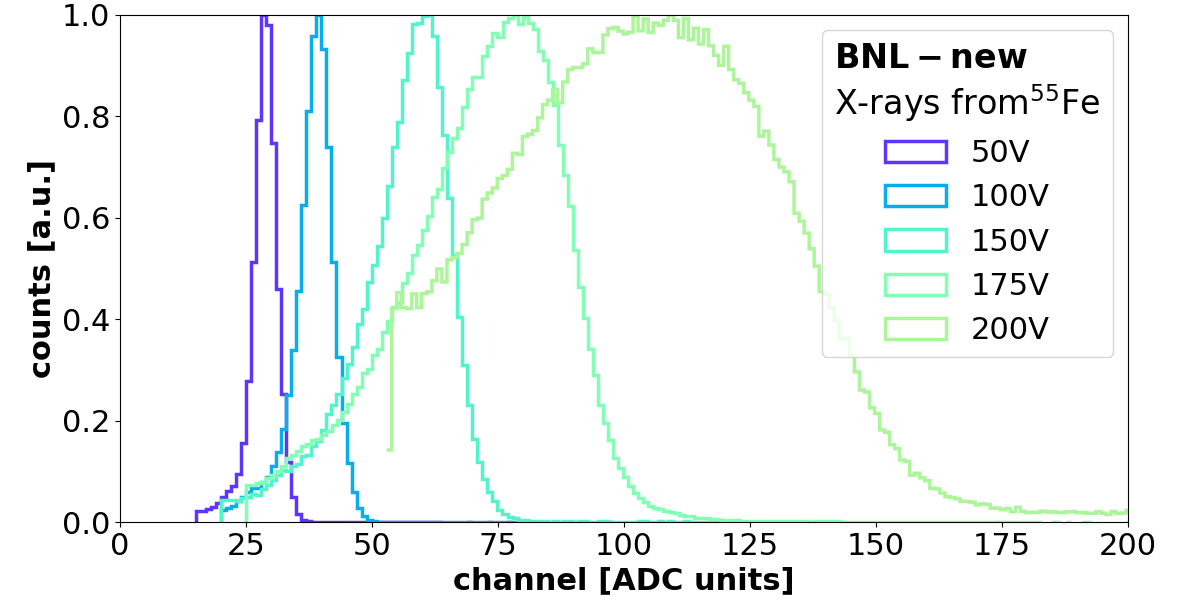}
	\caption{Pulse height spectra acquired by the BNL-new LGAD when exposed to, from top to bottom, beta from ${}^{90}$Sr, alphas from ${}^{241}$Am, and X-rays from ${}^{55}$Fe. Counts are normalized for better comparison. The charge on the \textit{x}-axis is represented in units of ADC "channels". The distributions are cut at lower ADC values to remove the contribution from the sensor noise.}
	\label{fig:BNLnew}
\end{figure}

\begin{figure}[!htbp]
    \centering\includegraphics[width=0.8\linewidth]{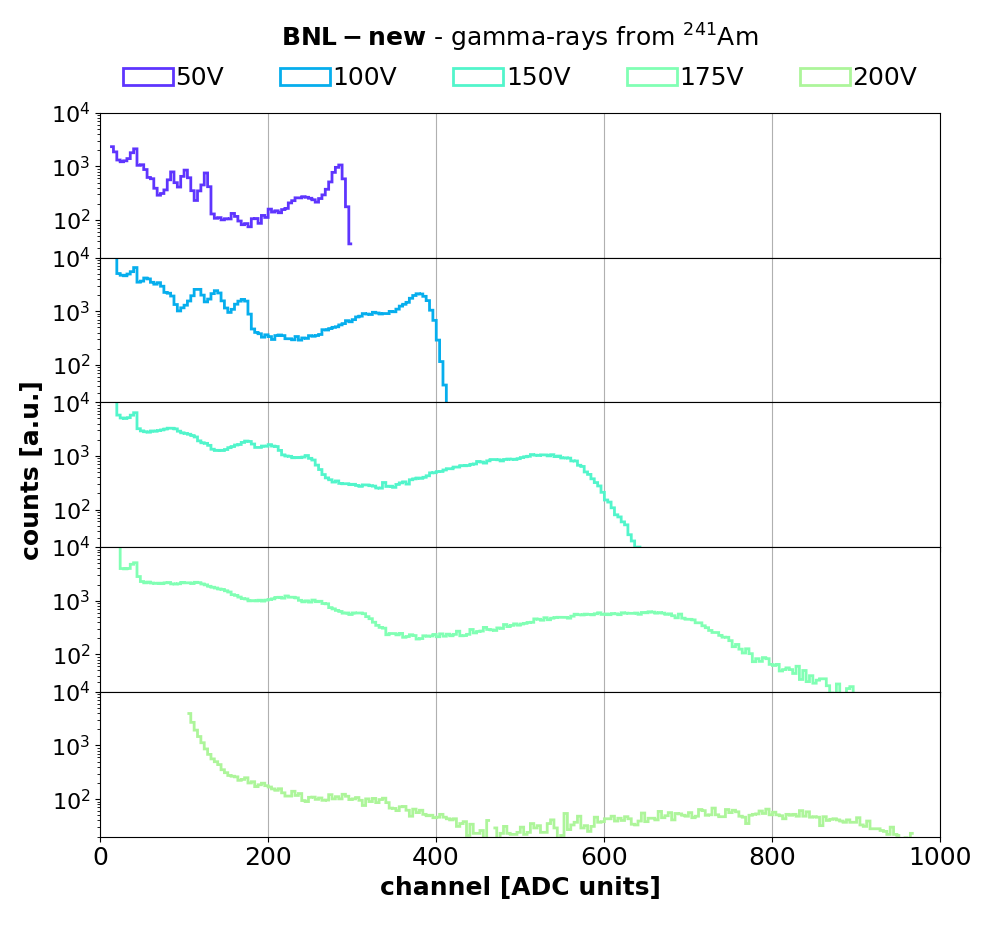}
    
	\caption{Pulse height spectra acquired by a BNL-new LGAD when exposed to gamma-rays from ${}^{241}$Am. Counts are normalized for better comparison. The charge on the \textit{x}-axis is represented in units of ADC "channels". The distributions are cut at lower ADC values to remove the contribution from the sensor noise.}
	\label{fig:BNLnew-2}
\end{figure}

In Fig.~\ref{fig:BNLnew} and ~\ref{fig:BNLnew-2} the spectra from the same sources as in Fig.~\ref{fig:BNLold} and ~\ref{fig:BNLold-2} are shown for the \textsc{BNL-new} LGAD, fabricated with a deeper gain layer. Voltages range from 50~V (just greater than the depletion voltage) to 200~V (close to the breakdown voltage). In the only case of  acquisitions with alpha, an additional   spectrum has been acquired for $V_{bias}=30~V$ as noticeably the gain is measured to be larger at this voltage. Better noise performance are clearly noticeable by looking at the low energy peaks of the  ${}^{241}$Am gamma-rays which, differing from the case of the \textsc{BNL-old} LGAD, are well resolved for the lowest voltages. Noticeably, the spectra from ${}^{55}$Fe are also visible in this device.

\begin{figure}[!htbp]
    \centering\includegraphics[width=.8\linewidth]{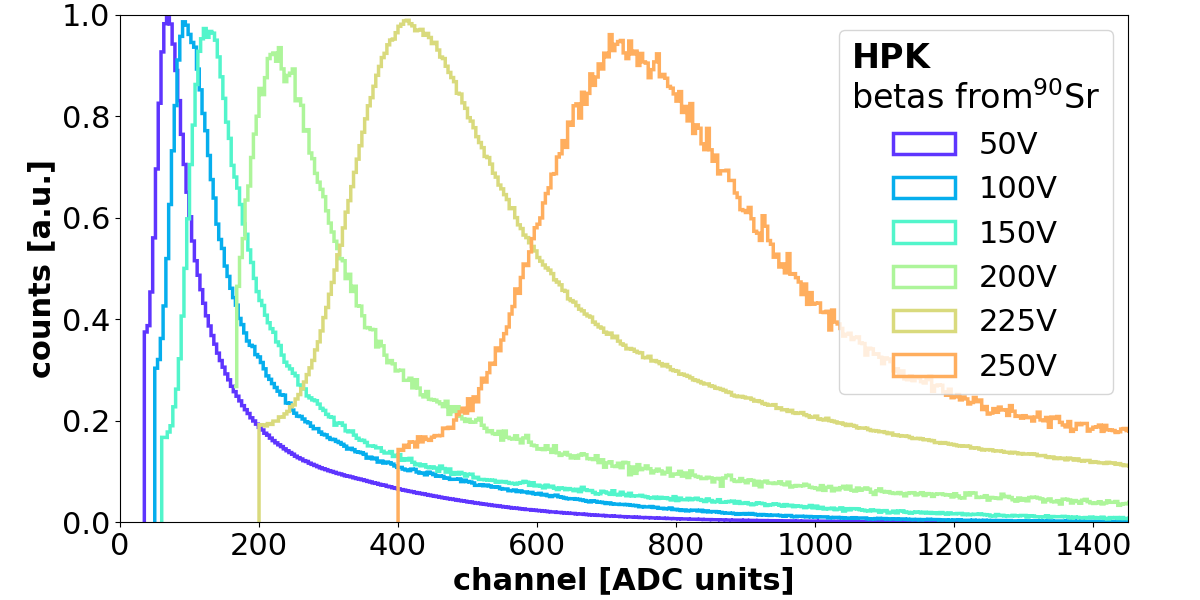}
    \centering\includegraphics[width=.8\linewidth]{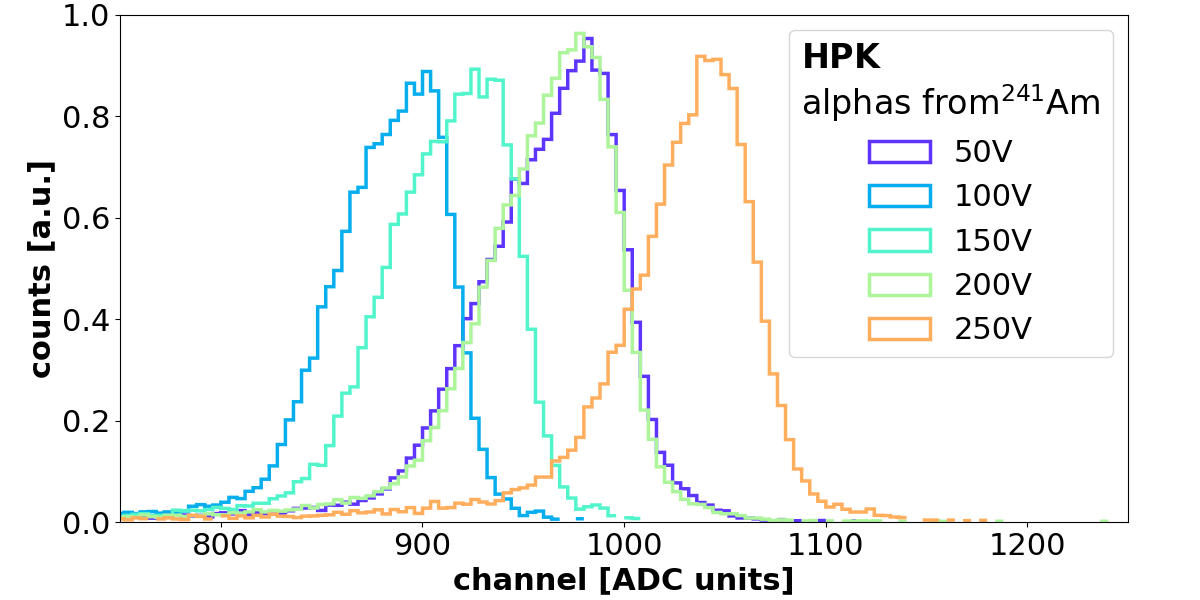}
    \centering\includegraphics[width=.8\linewidth]{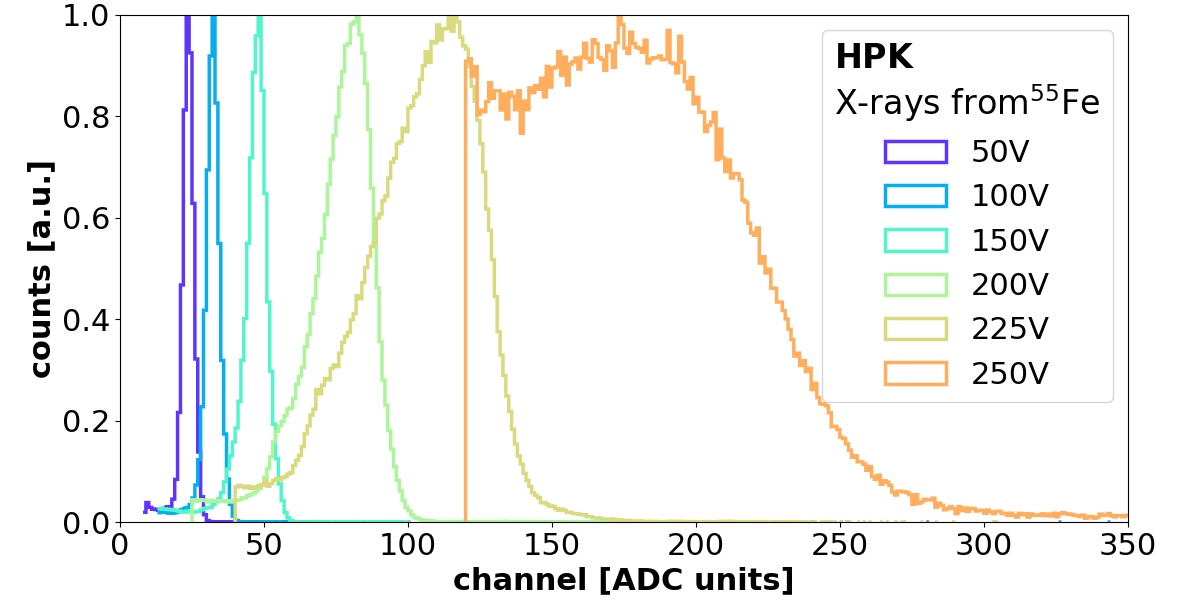}    
	\caption{Pulse height spectra acquired by the HPK LGAD when exposed to (from top to bottom) beta from ${}^{90}$Sr, alphas from ${}^{241}$Am, and X-rays from ${}^{55}$Fe. Counts are normalized for better comparison. The charge on the \textit{x} axis is represented in units of ADC "channels". The distributions are cut at lower ADC values to remove the contribution from the sensor noise.}
	\label{fig:HPK}
\end{figure}

\begin{figure}[!htbp]
    \centering\includegraphics[width=0.8\linewidth]{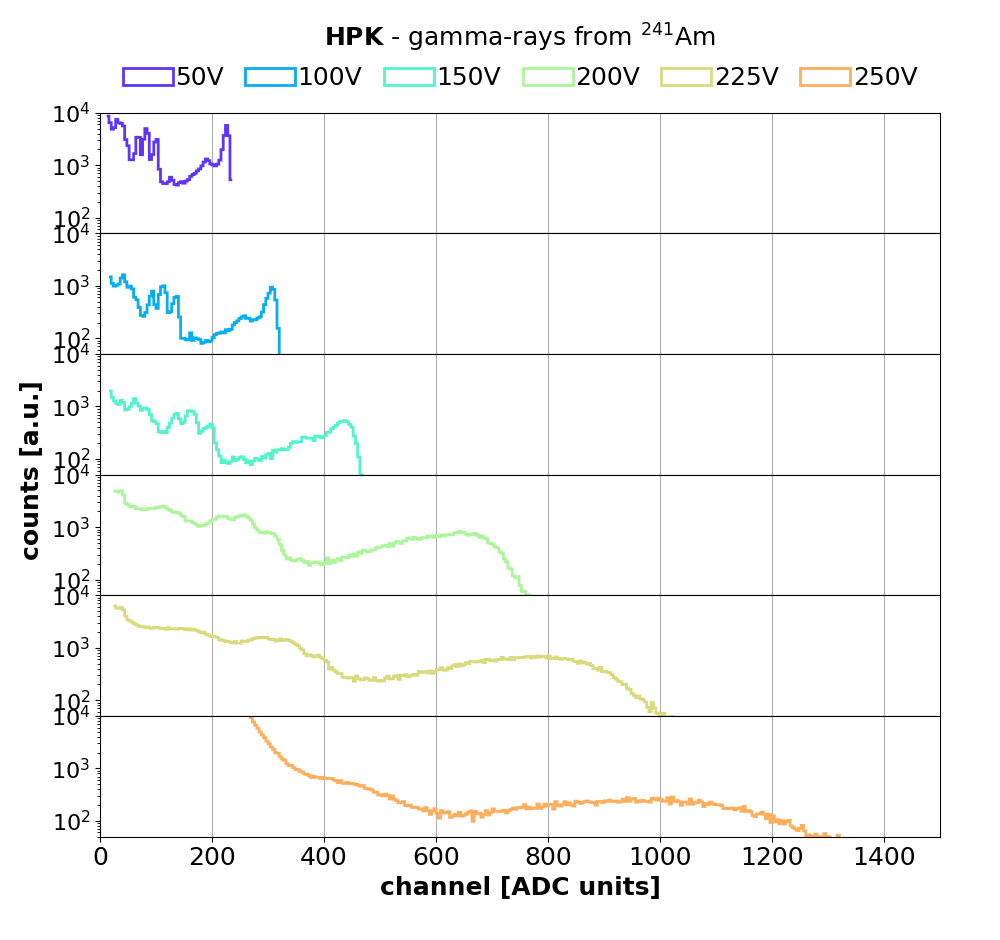}
    
	\caption{Pulse height spectra acquired by the HPK LGAD when exposed to gamma-rays from ${}^{241}$Am. Counts are normalized for better comparison. The charge on the \textit{x}-axis is represented in units of ADC "channels". The distributions are cut at lower ADC values to remove the contribution from the sensor noise.}
	\label{fig:HPK-2}
\end{figure}

In Fig.~\ref{fig:HPK} and ~\ref{fig:HPK-2}, the spectra acquired  with the \textsc{HPK} LGAD are reported for voltages ranging from 50~V (about depletion voltage) to 260~V (close to breakdown voltage).
Spectra for the HPK device are very similar to the ones obtained by the \textsc{BNL-new} sensor, except that larger gains are observed.

\subsection{Gain}

\begin{figure}[!htbp]
    \centering\includegraphics[width=0.49\linewidth]{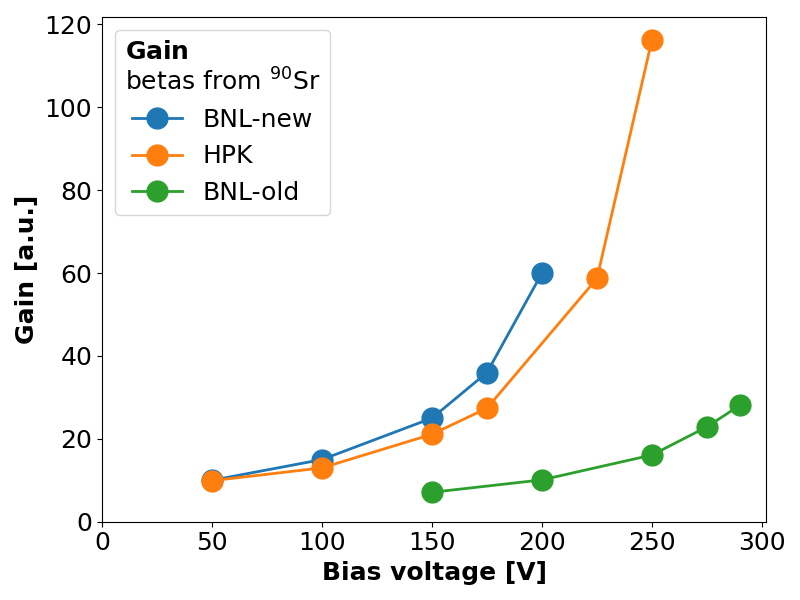}
    \centering\includegraphics[width=0.49\linewidth]{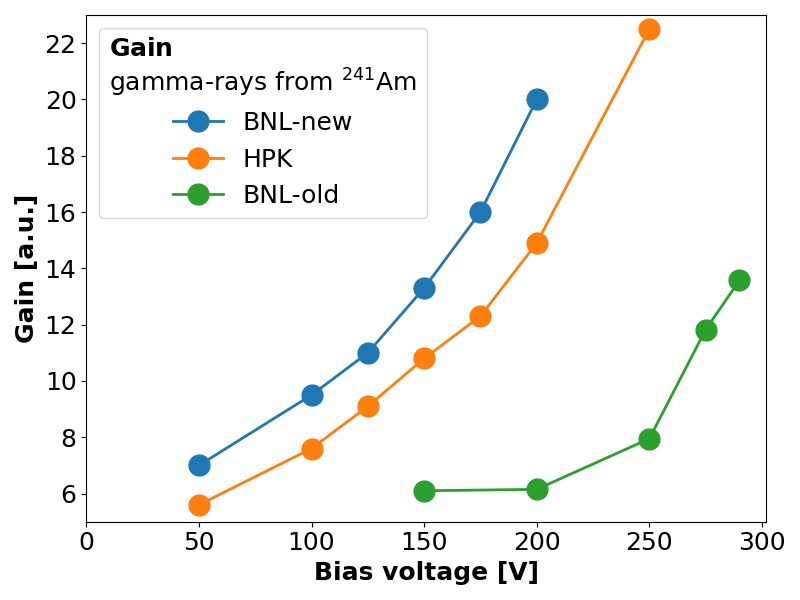}
    \centering\includegraphics[width=0.49\linewidth]{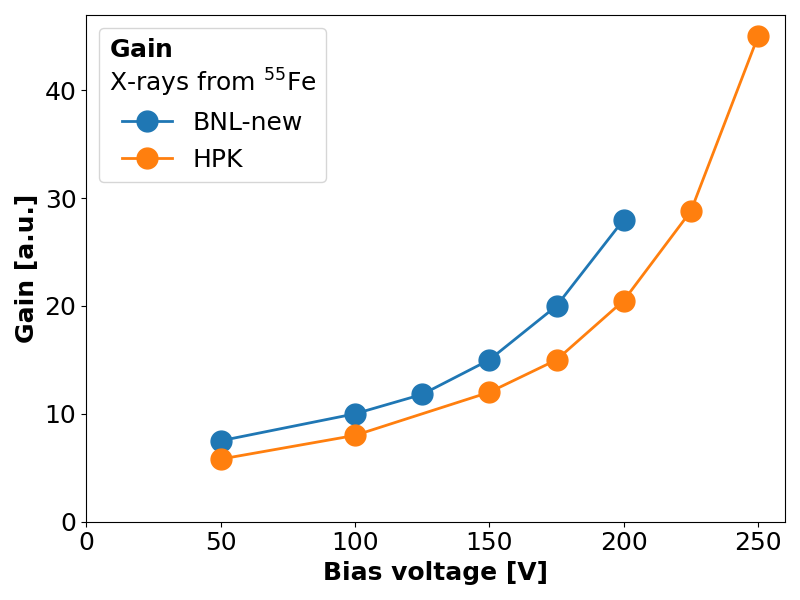}
    \centering\includegraphics[width=0.49\linewidth]{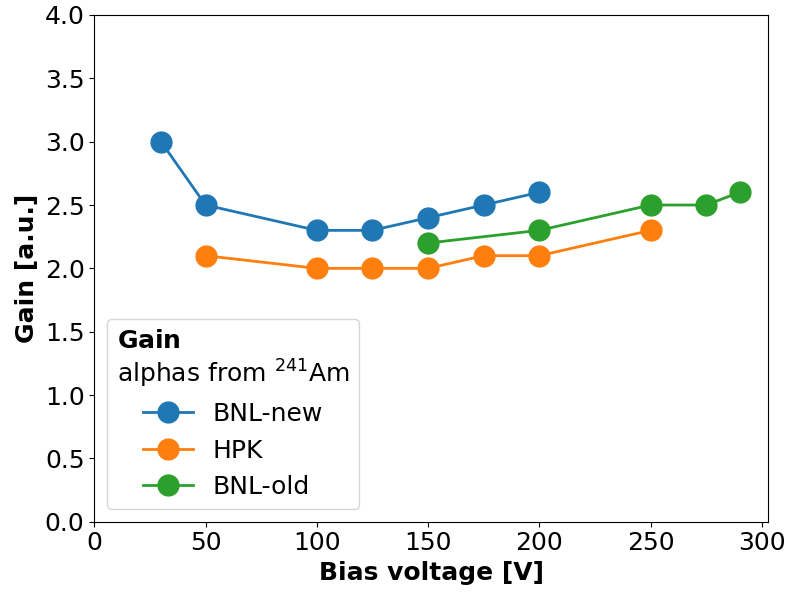}
	\caption{Summary of the average gain for the three devices under test as a function of bias voltage, when the devices are exposed to different radioactive sources.}
	\label{fig:gain}
\end{figure}

The summary of the average gain measured on the three LGADs under test is plotted in Fig.~\ref{fig:gain}; the gain as a function of the applied bias voltage is plotted for the different radioactive sources used for the evaluation. Gain values are obtained by comparing the signal produced by the LGAD sensors to that produced by a PIN diode exposed to the same radioactive sources and connected to the same readout system. 

For all devices under test, LGADs exhibit the largest gain when detecting betas from ${}^{90}$Sr. The largest gain can be attributed to the lowest density in the charge track of electron-hole pairs created by the impinging particles. Charges are distributed along a track rather than in a localized spot into the silicon bulk. The lowest gain (about 2 for all LGADs) is observed in the interaction with alpha particles, as a large number of electron-hole pairs is generated locally. For alphas and X- or gamma-rays, inner carriers in the charge cloud are shielded from the external electric field by the outermost charges, which furthermore lowers the local magnitude of the electric field. As a result, the electrons experience a limited multiplication.

Overall, for mips (as for beta particles from ${}^{90}$Sr) the maximum measured gain has a broad range and is about 120, 60, and 30 for the \textsc{HPK}, \textsc{BNL-new}, and \textsc{BNL-old} devices, respectively. Such a large variation is attributed to the different depth of the gain layer, with the \textsc{HPK} sensor having the deepest and \textsc{BNL-old} the shallowest.
When detecting X-rays, LGADs feature roughly half the gain observed for mips while they are operated in the same conditions. Larger gains are observed in the detection of low energy X-rays than in the detection of higher energy gamma-rays,  as a result of the lower carrier density of the former.

Particularly interesting is the case of the gain for alphas. For \textsc{HPK} and \textsc{BNL-new} LGADs, the gain is not a monotonic function of the bias voltage.  At the lowest voltages the LGADs show maximum gain, which then decreases when increasing the applied voltage only to eventually slightly increase again at higher voltage. This can be explained by the low electric field present in the bulk at low voltage, which causes a slow electron drift towards the multiplication region. This effect allows electrons to laterally diffuse and therefore a lower charge density is presented in the gain region~\cite{KrambergerFrontiers}. In this case the self-screening effect is therefore slightly mitigated when compared to faster drifts occurring at higher voltages. Finally, at the highest voltages the higher electric fields are responsible for an increase in gain, which compensates this effect.

\subsection{Signal-to-Noise Ratio}

\begin{figure}[!htbp]
    \centering\includegraphics[width=.8\linewidth]{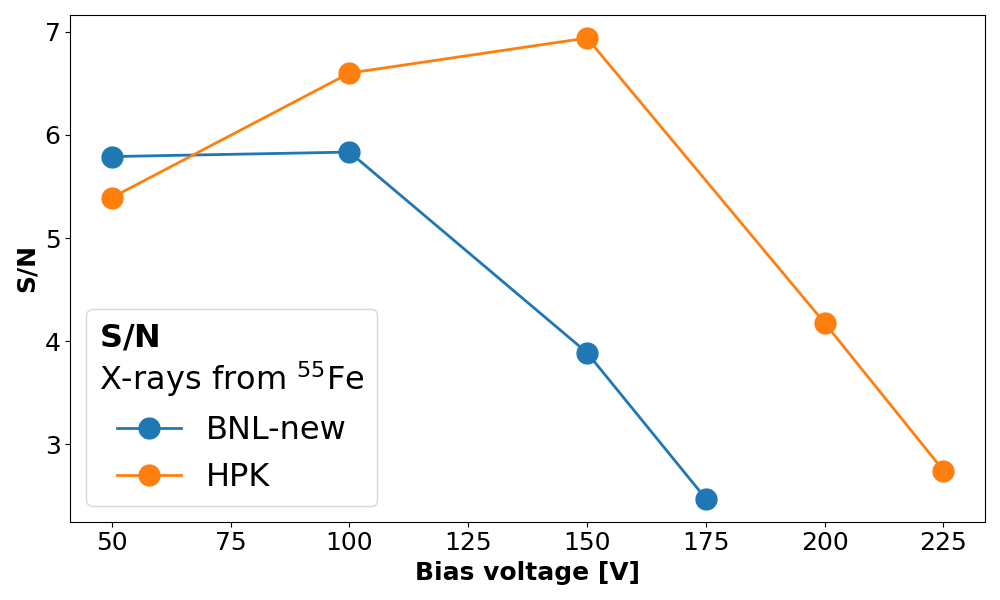}
    \centering\includegraphics[width=.8\linewidth]{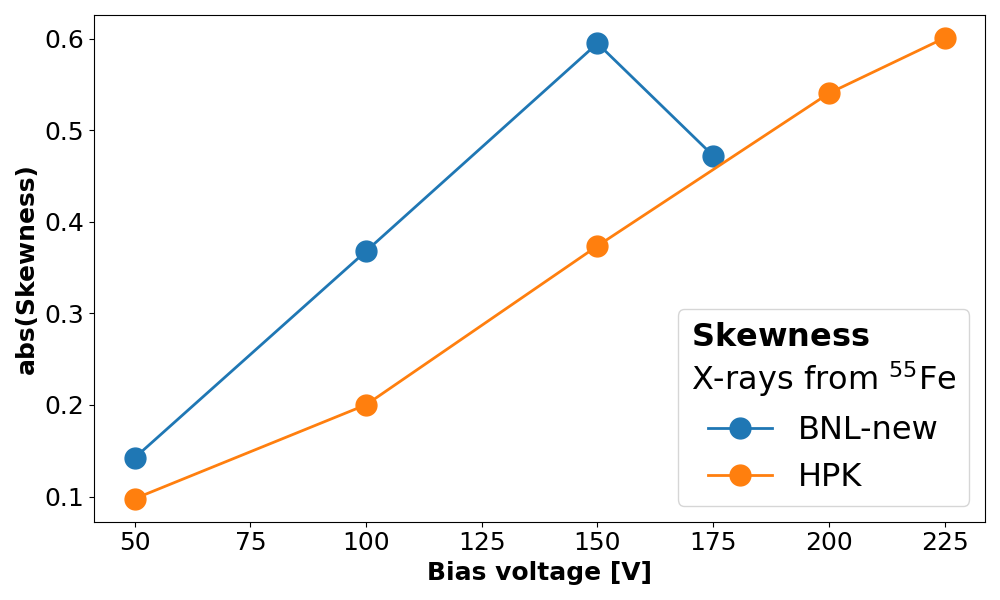}    
   	\caption{{\it Top}, signal-to-noise ratio (S/N)  for \textsc{BNL-new} and \textsc{HPK} LGADs and, 
    {\it Bottom}, the calculated skewness (in absolute value) as functions of the bias voltage using the ${}^{55}$Fe source.}
	\label{fig:SignalToNoise}
\end{figure}

\begin{figure}[]
    \centering\includegraphics[width=.8\linewidth]{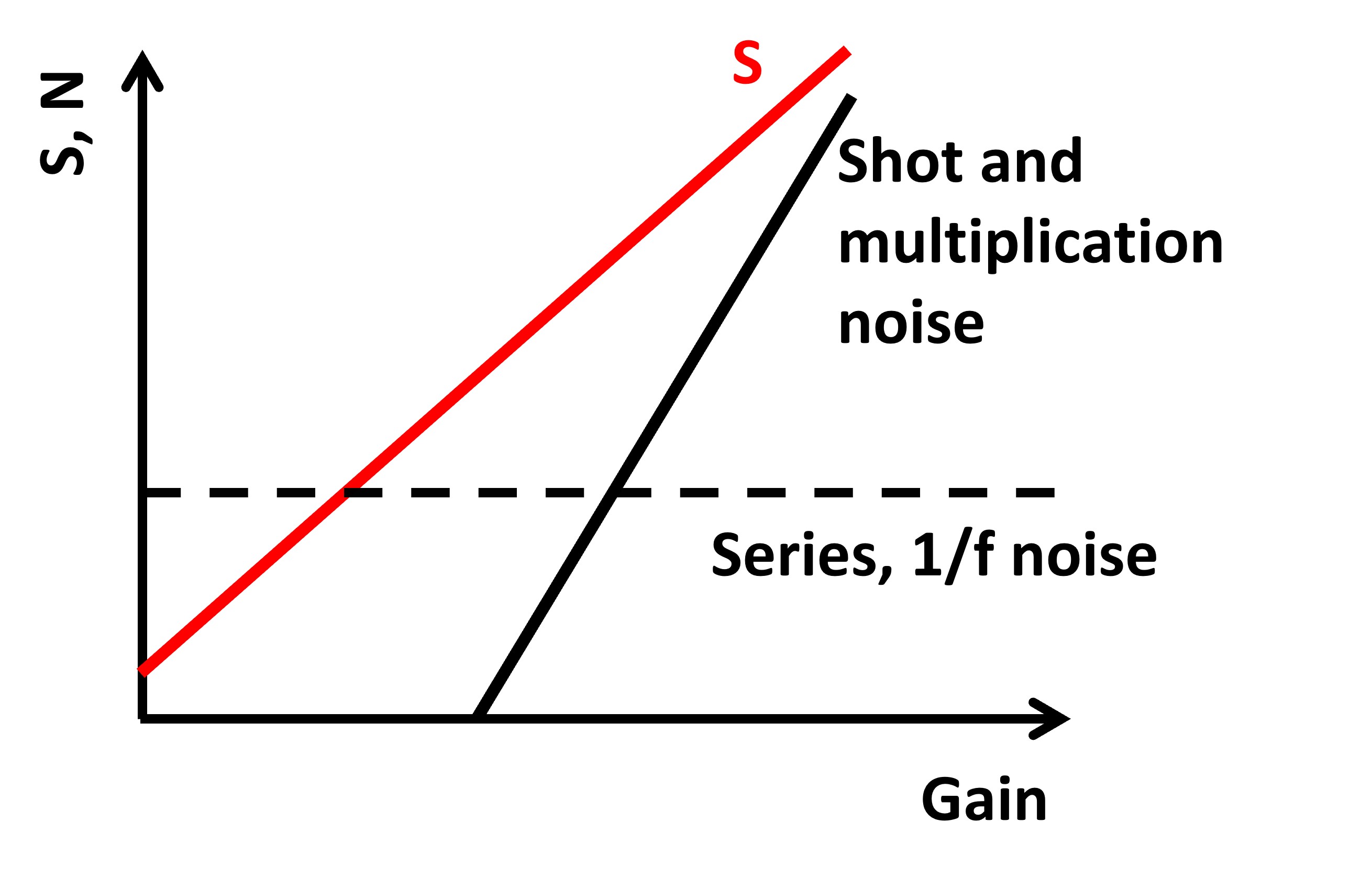}
       	\caption{Sketch of the signal (S) and the several noise sources (N) in an APD~\cite{Cartiglianoise} as a function of the gain.There is an optimal gain that maximize the S/N.}
	\label{fig:noise}
\end{figure}

The signal-to-noise ratio (S/N), defined as ratio between the distribution mode and its full width at half maximum (FWHM) and calculated using the ${}^{55}$Fe spectra acquired for the \textsc{BNL-new} and the \textsc{HPK} devices, is shown in Fig.~\ref{fig:SignalToNoise}. For the extraction of this quantity, we evaluated only the spectra from the ${}^{55}$Fe source as it creates in silicon only one narrow  line from its $K_{\alpha}$ decay at 5.9~keV (neglecting the less active  $K_{\beta}$ line at 6.4~keV):  it is therefore straightforward  to measure the S/N.  In the lowest bias voltage range, the S/N  slightly increases for increasing voltages until a maximum is reached, then it rapidly decreases at higher voltages, as the gain increases.
This can be better understood with the help of Fig.~\ref{fig:noise}~\cite{Cartiglianoise}, showing the signal and the noise components as a function of the gain for a generic avalanche photodiode. It shows how a gain increase brings the signal over the noise threshold in the detector. The series and 1/f electronic noise sources of the system are independent of the gain, while the shot noise and the multiplication~\cite{McIntyre_noise} noises increase more than linearly with the gain. There exists therefore an optimal gain that maximizes the S/N while higher gains are in fact detrimental as far as S/N is concerned.

The spectra are not symmetric around the mode but rather they show a skewness in the lower energy shoulder. This effect suggests that the multiplication process is biased towards smaller gain values. The skewness for the same ${}^{55}$Fe datasets is calculated as the Pearson's first skewness coefficient:
\begin{equation}
    \rm{skewness = \frac{mean-mode}{sigma}}
\end{equation}
where sigma is the standard deviation of the pulse height spectra; the skewness of the dataset as a function of the bias voltage is also shown in Fig.~\ref{fig:SignalToNoise}. The skewness is observed to rapidly increase as the bias voltage increases, albeit a minor divergence from this trend can be observed in the \textsc{BNL-new} dataset for high bias voltages due to the high noise of the sensor. It can also be noticed that even at the smaller voltages (close to depletion) the skewness differs from zero.

\section{Conclusions}
By acquiring the pulse height spectra with a variety of LGADs exposed to different radioactive sources and at different bias voltages, we were able to characterize gain and noise properties.  The gain of LGAD sensors depends on the nature of the detected particle, being maximum for mips, which create a low density of electron-hole pairs along a particle track, and minimum for alpha particles that create a large number of electron-hole pairs in a localized cloud. Moreover, the gain changes drastically as a function of the depth of the gain layer, being higher in sensors fabricated with deeper gain implants. 
The spectroscopic performance, quantified by the S/N ratio, rapidly deteriorates by increasing the bias voltage and thus the gain, as the multiplication and shot noises dominate over the other noise sources and overcompensates for the signal gain. It appears therefore that there is an optimal gain, usually small, that maximize spectroscopic performance. We also observed how another spectroscopic parameter,  the skewness, increases as a function of bias voltage as a lower energy shoulder is featured in the spectra, indicating a  tendency of the multiplication process towards smaller than average  amplification values.


\bibliographystyle{elsarticle-num}
\bibliography{sample.bib}








\end{document}